\documentclass[12pt]{article}

\usepackage[english]{babel}
\usepackage{latexsym}

\begin{document}

\title{1D quantum models with correlated disorder vs. classical
oscillators with coloured noise}

\author{L.~Tessieri \\
{\it Department of Chemistry, Simon Fraser University, Burnaby,} \\
{\it British Columbia, Canada V5A 1S6} \\
F.~M.~Izrailev \\
{\it Instituto de F\'{\i}sica, Universidad Aut\'{o}noma de Puebla,} \\
{\it Apartado Postal J-48, Puebla, 72570, Mexico}}

\date{4th June 2001}

\maketitle

\begin{abstract} 
We perform an analytical study of the correspondence between a
classical oscillator with frequency perturbed by a coloured noise
and the one-dimensional Anderson-type model with correlated diagonal
disorder.
It is rigorously shown that localisation of electronic states in the
quantum model corresponds to exponential divergence of nearby
trajectories of the classical random oscillator.
We discuss the relation between the localisation length for the
quantum model and the rate of energy growth for the stochastic
oscillator.
Finally, we examine the problem of electron transmission through a
finite disordered barrier by considering the evolution of the classical
oscillator.
\end{abstract}

{Pacs numbers: 05.40.-a, 71.23.An, 72.15.Rn}

\section{Introduction}

This work serves the goal of establishing some quantitative links
between two seemingly unrelated fields: quantum disordered models
on the one hand and classical stochastic systems on the other.
More precisely, we analyse the relations existing between a classical
oscillator with frequency perturbed by a feeble noise and the
one-dimensional (1D) Anderson-type model with a weak diagonal
disorder. Our main interest is in {\em correlated} random potentials
and, correspondingly, in {\em coloured} noise for the stochastic
oscillator.

Recently, the r\^{o}le of correlations in random potentials of
quantum models has been the object of intense scrutiny.
In particular, it was shown that specific long-range correlations
in potentials may lead to the emergence of a continuum of extended
states even in 1D lattices (see, e.g.,~\cite{Izr99, Kuh00} and
references therein). In this paper we show that the phenomenon of
Anderson localisation has its counterpart in the energetic instability
of a random oscillator. Specifically, the mobility edge generated
in the 1D quantum models by long-range correlations is equivalent
to the suppression of the energy growth of the stochastic oscillator
due to temporal correlations of the frequency noise.

We use the correspondence between stochastic oscillators and
disordered solid state models in order to study the transmission
properties of finite lattices by making use of the dynamical analysis
of an oscillator with noisy frequency.
This approach allows us to put in a new perspective the problem of
electronic transport in disordered lattices and also to gain new
insight on the dynamics of random oscillators.

This paper is organised as follows. In the following section we
define the models that constitute the object of our study, and we make
some general considerations on their analogies. In Sec.~\ref{lyasec}
we rigorously analyse the relation between the localisation of electronic
states for the Anderson model and the orbit instability of a random
oscillator.
In Sec.~\ref{mobsec} we discuss how correlations of the frequency
noise can suppress the energy growth of the stochastic oscillator.
The analogy between a random oscillator and a disordered chain is then
used in Sec.~\ref{trasec} to study the electronic transmission
through a finite disordered lattice. In Sec.~\ref{trasec} we also
discuss the relation between energetic instability and orbit divergence
for a random oscillator.
Finally, Sec.~\ref{concl} is devoted to the summarising conclusions.

\section{Definition of the models}
\label{defmod}

The Anderson model is defined by the discrete stationary
Schr\"{o}dinger equation
\begin{equation}
\psi_{n+1} + \psi_{n-1} + \varepsilon_{n} \psi_{n} = E \psi_{n}
\label{andmod}
\end{equation}
where $\psi_{n}$ is the amplitude of the wave function at the $n$th
site of the lattice and disorder is introduced via the site energies
$\varepsilon_{n}$ which are assumed to be random correlated variables.
We do not restrict our considerations to a specific distribution for
the random potential $\varepsilon_{n}$; we only suppose that it has zero
average $\langle \varepsilon_{n} \rangle = 0$ and that the binary
correlator $\langle \varepsilon_{n} \varepsilon_{n+k} \rangle$
is a known function of the index $k$. We also assume that the correlator
$\langle \varepsilon_{n} \varepsilon_{n+k} \rangle$ does
not depend on $n$ and that it is a decreasing function of $k$.
In other words, we make the physically sensible assumptions that the
random succession $\{ \varepsilon_{n} \}$ is stationary, and that
correlations decay with increasing distance.
We restrict our analysis to the case of weak disorder, defined by the
condition
\begin{displaymath}
\langle \varepsilon_{n}^{2} \rangle \ll 1 .
\end{displaymath}
In the preceding expressions the symbol $\langle \ldots \rangle$ stands
for the average over a single disorder realization defined by the limit
\begin{displaymath}
\langle x_{n} \rangle = \lim_{N \rightarrow \infty} \frac{1}{N}
\sum_{n=1}^{N} x_{n} ;
\end{displaymath}
we assume that this average is equivalent to the average over disorder
realizations (ensemble average) for the succession $\{ \varepsilon_{n} \}$.

It is known that the model~(\ref{andmod}) can be put into
correspondence with the kicked oscillator defined by the
Hamiltonian
\begin{equation}
H = \omega \left( \frac{x^2}{2} + \frac{p^2}{2} \right) +
\frac{x^2}{2} \left( \sum_{n=-\infty}^{\infty} A_{n} \delta(t-nT)
\right),
\label{kickos}
\end{equation}
which represents an oscillator whose momentum undergoes instantaneous
variations of random intensity $A_{n}$ at regular time intervals.
The connection between the models~(\ref{andmod}) and~(\ref{kickos})
has been discussed before (see, e.g.,~\cite{Izr95}). Basically, the
correspondence consists in the fact that, by integrating the
Hamilton equations of motion of the oscillator~(\ref{kickos}) over the
period between two successive kicks one gets the map
\begin{equation}
\begin{array}{ccc}
x_{n+1} & = & x_{n} \cos (\omega T) + (p_{n} - A_{n} x_{n})
\sin (\omega T) \\
p_{n+1} & = & -x_{n} \sin (\omega T) + (p_{n} - A_{n} x_{n})
\cos (\omega T)
\end{array}
\label{map}
\end{equation}
where $x_{n}$ and $p_{n}$ stand for the position and momentum of the
oscillator immediately before the $n$th kick.
This map is equivalent to to the Schr\"{o}dinger equation~(\ref{andmod})
which defines the Anderson model. Indeed, by eliminating the momentum
from Eqs.~(\ref{map}), one gets the relation
\begin{displaymath}
x_{n+1} + x_{n-1} + A_{n} \sin (\omega T) x_{n} = 2x_{n}
\cos (\omega T)
\end{displaymath}
which coincides with the Schr\"{o}dinger equation~(\ref{andmod})
provided that the position $x_{n}$ of the oscillator at time $t=nT$
is identified with the electron amplitude $\psi_{n}$ at the $n$th site
and that the parameters of the kicked oscillator are related to those
of the Anderson model by the identities
\begin{equation}
\begin{array}{ccc}
\varepsilon_{n} = A_{n} \sin (\omega T) & \mbox{and} & E = 2 \cos
(\omega T).
\end{array}
\label{corres}
\end{equation}

The formal correspondence between the quantum model~(\ref{andmod})
and the kicked oscillator~(\ref{kickos}) raises the question of
whether a similar analogy can link the Anderson model to a
random oscillator whose frequency is perturbed by a continuous
noise instead that by a succession of discontinuous and singular
kicks as in model (\ref{kickos}). In other words, one is led
to infer the existence of close ties between the quantum
model~(\ref{andmod}) and a stochastic oscillator defined by the
Hamiltonian
\begin{equation}
H = \omega \left( \frac{x^2}{2} + \frac{p^2}{2} \right) +
\frac{x^2}{2} \xi(t)
\label{randosc}
\end{equation}
where $\xi(t)$ is a continuous and stationary noise.
Notice that these requirements on $\xi(t)$ set the random
oscillator~(\ref{randosc}) and the kicked oscillator~(\ref{kickos}) in
two different categories within the vast family of stochastic oscillators,
since the succession of kicks in the model~(\ref{kickos}) is a
non-stationary and strongly discontinuous random process.
Consequently, the connection between the Anderson model~(\ref{andmod})
and the kicked oscillator~(\ref{kickos}) does not prove at all the
equivalence of models~(\ref{andmod}) and~(\ref{randosc}) but only
constitute a hint that such a link may exist.

In our analysis of stochastic oscillators, we will focus on the
Hamiltonians represented by Eq.~(\ref{randosc}), completing the
definition of the model by further assuming that the noise
$\xi(t)$ has zero average and that its binary correlator is a
known function
\begin{equation}
\begin{array}{ccc}
\langle \xi(t) \rangle = 0 & \mbox{and} &
\langle \xi(t) \xi(t+\tau) \rangle = \chi (\tau) .
\end{array}
\label{noisprop}
\end{equation}
In Eq.~(\ref{noisprop}) the symbol $\langle \ldots \rangle$ is used for
the time average
\begin{displaymath}
\langle f(t) \rangle = \lim_{T_{0} \rightarrow \infty} \frac{1}{T_{0}}
\int_{0}^{T_{0}} f(t) \; dt ,
\end{displaymath}
which is assumed to coincide with the ensemble average for the process
$\xi(t)$.
Notice that we do not restrict our consideration to the case of
white noise, but we are instead interested in the general case of
{\em coloured} noise.
Finally, we require that the noise $\xi(t)$ be weak; in other words,
we assume that the fluctuations of the frequency around its average
value are small.

Below we show that oscillators of the kind~(\ref{randosc}), with
the above-mentioned noise features, are equivalent to the Anderson
model~(\ref{andmod}) if two further conditions are met.
First, the correlation function has to be of the form
\begin{equation}
\chi (\tau) = \frac{\langle A_{n}^{2} \rangle}{T}
\sum_{k=-\infty}^{+\infty} \zeta (k) \; \delta (\tau - k T) ,
\label{corfun}
\end{equation}
where the symbol $\zeta(k)$ stands for the normalised binary correlators
\begin{equation}
\zeta(k) = \frac{ \langle A_{n+k} A_{n} \rangle}{\langle A_{n}^{2}
\rangle}
\label{bincor}
\end{equation}
of the random variables $A_{n}$ specified by the second condition.
Our second requirement is
that the unperturbed frequency $\omega$ of the oscillator and the
parameters $A_{n}$ must be related to the parameters $E$ and
$\varepsilon_{n}$ of the Anderson model through the
identities~(\ref{corres}).

Notice that the links established by these two conditions associate
key features of the noise $\xi(t)$ to the corresponding properties
of the random potential $\varepsilon_{n}$. Indeed, once the random
variables $\varepsilon_{n}$ and $A_{n}$ are connected by the
relation~(\ref{corres}), the correlators~(\ref{bincor}) become identical
to the normalised correlators of the potential $\varepsilon_{n}$.
Therefore the spatial correlations of the disorder in the Anderson model
are mirrored by temporal correlations for the noise $\xi(t)$.
In the special case in which the disorder in the Anderson model is
{\em uncorrelated} (i.e., $\langle \varepsilon_{n+k} \varepsilon_{n}
\rangle = 0$ for $k \ne 0$), the noise for the random oscillator is
{\em white} (i.e., $\langle \xi(t) \xi(t+\tau) \rangle \propto
\delta(\tau)$).
One can also observe that the case of weak disorder in the Anderson
model corresponds to that of weak noise for the random oscillator, since
the condition $\langle \varepsilon_{n}^{2} \rangle \ll 1$ entails the
consequence that $\langle A_{n}^{2} \rangle \ll 1$ (except that at the
band edge, i.e., for $\omega T \rightarrow 0$, which is a special case
where anomalies are expected to arise and will not be considered here).

Obviously, we must endow with a well-defined meaning the notion of
``equivalence'' used above to describe the connection between
the Anderson model~(\ref{andmod}) and the random
oscillator defined by Eqs.~(\ref{randosc}) and~(\ref{corfun}).
We speak of equivalence of the two models in the sense that the
time evolution of the orbits of the random oscillator closely mirrors
the spatial variation of the electronic states on the lattice.
More precisely, the exponential divergence rate of nearby orbits
turns out to be equal to the inverse localisation length of the
Anderson model.

The correspondence between the random oscillator~(\ref{randosc}) and
the Anderson model~(\ref{andmod}) is to some extent surprising since
the former is a {\em classical} system and is {\em continuous} in time
whereas the latter model is {\em quantum} and {\em discrete} in space.
It is therefore particularly interesting to notice how close the two
systems turn out to be.
To sum up, one of the main results of this paper is that the Anderson
model with weak {\em correlated} disorder has a close analogue in a
random oscillator with frequency perturbed by a {\em coloured} noise.
This equivalence generalises the result established in Ref.~\cite{Tes00}
where the Anderson model with uncorrelated disorder was linked to
a random oscillator of the kind~(\ref{randosc}) with white noise.

\section{The Lyapunov exponent}
\label{lyasec}

In the previous section we have described the analogy between the
Anderson model~~(\ref{andmod}) and the random oscillator~(\ref{randosc})
as being based on the correspondence between the electronic
wave-function of the former model and the space orbits of the latter
system.
To prove this analogy, we will compute the divergence rate of nearby
trajectories of the random oscillator, i.e., its Lyapunov exponent and
we will show that, when the conditions~(\ref{corres}) and~(\ref{corfun})
are met, the Lyapunov exponent coincides with the inverse localisation
length in the Anderson model.
We define the Lyapunov exponent through the formula
\begin{equation}
\lambda = \lim_{T_{0} \rightarrow \infty} \lim_{\delta \rightarrow 0}
\frac{1}{T_{0}} \frac{1}{\delta} \int_{0}^{T_{0}}
\ln \frac{x(t+\delta)}{x(t)} \; dt .
\label{lyap}
\end{equation}

To compute this expression it is convenient to introduce the
polar coordinates defined through the standard relations $x = r
\sin \theta$, $p = r \cos \theta$. This allows one to cast
Eq.~(\ref{lyap}) in the form
\begin{displaymath}
\lambda = \lim_{T_{0} \rightarrow \infty} \frac{1}{T_{0}} \int_{0}^{T_{0}}
\frac{\dot{r}}{r} \; dt .
\end{displaymath}
To proceed further, we consider the dynamical equations for the
random oscillator in polar coordinates
\begin{equation}
\dot{\theta} = \omega + \xi(t) \sin^{2} \theta ,
\label{ang}
\end{equation}
\begin{equation}
\dot{r} = - \frac{1}{2} r \xi(t) \sin 2 \theta ;
\label{rad}
\end{equation}
Using the radial Eq.~(\ref{rad}), the expression for the Lyapunov
exponent can be finally put into the form
\begin{equation}
\lambda = - \lim_{T_{0} \rightarrow \infty} \frac{1}{2T_{0}}
\int_{0}^{T_{0}}
\xi(t) \sin \left( 2 \theta(t) \right) \; dt = -\frac{1}{2} \langle
\xi(t) \sin \left( 2 \theta(t) \right) \rangle .
\label{lyap2}
\end{equation}

The problem of computing the Lyapunov exponent~(\ref{lyap}) is thus
reduced to that of calculating the noise-angle correlator that appears
in Eq.~(\ref{lyap2}).
This can be done in the following way, which is the extension to
the continuum case of the procedure adopted in~\cite{Izr99} for
the discrete case.
First, one introduces the noise-angle correlator defined by the
relation
\begin{displaymath}
\gamma(\tau) = \langle \xi(t) \exp \left( 2 i \theta \left( t +
\tau \right) \right) \rangle .
\end{displaymath}
Starting from this definition, in the limit $\epsilon \rightarrow 0$
one has
\begin{displaymath}
\gamma ( \tau + \epsilon ) =  \langle \xi(t) \exp \left( i 2 \theta
\left( t + \tau \right) \right) \left( 1 + 2 i \dot{\theta} \left(
t + \tau \right) \epsilon \right) \rangle + o (\epsilon) .
\end{displaymath}
With use of the dynamical equation~(\ref{ang}) one can further write
\begin{displaymath}
\begin{array}{ccl}
\gamma ( \tau + \epsilon ) & = & \gamma (\tau) \left( 1 + 2 i \omega
\epsilon \right) \\
 & & + 2 i \epsilon \langle \xi(t) \xi(t+\tau)
\exp \left( 2 i \theta \left( t+\tau \right) \right)
\sin^{2} \theta \left( t+\tau \right) \rangle + o (\epsilon) .
\end{array}
\end{displaymath}
In the limit of weak noise, one can factorise the correlator
that appears in the right hand side of the preceding equation and take
the average over the angular variable using a flat distribution for
$\theta$. Indeed, when  $\xi(t) \rightarrow 0$, Eq.~(\ref{ang}) implies
that $\dot{\theta} \simeq \omega$ so that, after a conveniently long
time, one can expect the angular variable to take values uniformly
distributed in the interval $[0,2 \pi]$.
As a consequence the noise-angle correlator must obey the
relation
\begin{equation}
\gamma ( \tau + \epsilon ) = \gamma (\tau) \left( 1 + 2 i \omega
\epsilon \right) - \frac{i}{2} \chi(\tau) \epsilon + o (\epsilon) ,
\label{gam1}
\end{equation}
where $\chi(\tau)$ is the correlation function (or noise-noise
correlator) defined by Eq.~(\ref{noisprop}).
On the other hand, a simple application of calculus rules leads to
\begin{equation}
\gamma ( \tau + \epsilon ) = \gamma (\tau) + \frac{d \gamma}{d \tau}
(\tau) \epsilon + o(\epsilon) .
\label{gam2}
\end{equation}
Comparing Eqs.~(\ref{gam1}) and~(\ref{gam2}), one obtains the
differential equation
\begin{displaymath}
\frac{d \gamma}{d \tau}(\tau) = 2 i \omega \gamma(\tau) -
\frac{i}{2} \chi(\tau) ,
\end{displaymath}
whose solution (with the boundary condition $\lim_{\tau \rightarrow
-\infty} \gamma (\tau) = 0$) gives the noise-angle correlator
\begin{displaymath}
\gamma (\tau) = - \frac{i}{2} \int_{-\infty}^{\tau} \chi(s)
e^{2 i \omega \left( \tau - s \right)} \; ds .
\end{displaymath}
Using this result the Lyapunov exponent~(\ref{lyap2}) can be
finally written as
\begin{equation}
\lambda = \frac{1}{8} \int_{-\infty}^{+\infty}
\langle \xi(t) \xi(t+\tau) \rangle \cos (2 \omega \tau) \; d \tau
\label{lyap3}
\end{equation}
which implies that the Lyapunov exponent for the stochastic
oscillator~(\ref{randosc}) is proportional to the Fourier transform
$\tilde{\chi}(2\omega)$ of the correlation function at twice the
frequency of the unperturbed oscillator.

We are interested in the particular case in which the correlation
function of the noise $\xi(t)$ takes the specific form~(\ref{corfun}),
because we want to prove that in that case the Lyapunov
exponent~(\ref{lyap3}) coincides with the inverse localisation length
of the Anderson model~(\ref{andmod}).
The substitution of the correlation function~(\ref{corfun}) in the
general expression~(\ref{lyap3}) gives
\begin{displaymath}
\lambda = \frac{\langle A_{n}^{2} \rangle}{8T}
\left[ 1 + 2 \sum_{k=1}^{+\infty} \zeta (k) \, \cos \left(
2 \omega T k \right) \right] .
\end{displaymath}
Taking also into account the relations~(\ref{corres}) between the
parameters of the systems~(\ref{andmod}) and~(\ref{randosc}), one can
finally write the Lyapunov exponent for the random oscillator as
\begin{equation}
\begin{array}{lcl}
\displaystyle
\lambda = \frac{1}{T} \frac{\langle \varepsilon_{n}^{2} \rangle}
{8 \sin^{2} \left( \omega T \right)} \; \varphi \left( \omega T \right)
& \mbox{with} &
\displaystyle
\varphi \left( \omega T \right) = 1 + 2
\sum_{k=1}^{+\infty} \zeta (k) \, \cos \left( 2 \omega T k \right).
\end{array}
\label{lyap4}
\end{equation}

This expression coincides with the one given in~\cite{Izr99} for the
localisation length in the Anderson model with correlated disorder.
The inverse localisation length is given by the product of two
factors, namely the Lyapunov exponent for the uncorrelated disorder
case and the function $\varphi (\omega T)$, which describes the
effect of disorder correlations (and which reduces to unity when
correlations are absent).
Formula~(\ref{lyap4}) thus confirms the equivalence of the quantum
Anderson model~(\ref{andmod}) with the classical
oscillator~(\ref{randosc}) which had been inferred in
Sec.~\ref{defmod} by the existence of a third system -the kicked
oscillator~(\ref{kickos})- which was somehow contiguous to both
model~(\ref{andmod}) and~(\ref{randosc}).
To sum up, formula~(\ref{lyap4}) allows one to conclude that the
Anderson model with {\em correlated} disorder has a classical counterpart
represented by a stochastic oscillator with frequency perturbed by a
{\em coloured} noise. This conclusion generalises the equivalence
established in~\cite{Tes00} between the Anderson model with {\em
uncorrelated} disorder and an oscillator with frequency perturbed by
a {\em white} noise.

A remark is in order here: expression~(\ref{lyap4}) for the inverse
localisation length of model~(\ref{andmod}) is correct for all energy
values inside the unperturbed band {\em except} that at the band
centre, i.e., for $\omega T = \pi/2$ where a anomaly arises and
special methods are required for the analytical investigation
(see, e.g.,~\cite{Kap81}). This anomaly is a resonance effect inherent
in the discrete nature of the model~(\ref{andmod}) and cannot
therefore be reproduced by the continuos system~(\ref{randosc}).
Other anomalies appear in the Anderson model for the ``rational''
values of the energy (i.e., when $\omega T = \pi p/2q$ with $p$ and
$q$ integer numbers), but they are effects of order higher than the
second~\cite{Tes00} and need therefore not be considered here.
In conclusion, apart from the exceptional case of the band centre,
the dynamical features of the models~(\ref{andmod})
and~(\ref{randosc}) do not differ to the second order of perturbation
theory.

The equivalence of the models~(\ref{andmod}) and~(\ref{randosc}) can
be examined also from a different point of view: that of the
correspondence between discrete and continuous solid state systems.
Indeed, the dynamical equation for the oscillator~(\ref{randosc})
\begin{equation}
\ddot{x} + \omega \xi(t) x = - \omega^{2} x
\label{osc}
\end{equation}
coincides, {\em mutatis mutandis}, with the stationary Schr\"{o}dinger
equation
\begin{equation}
- \psi'' + k \xi(x) \psi = k^{2} \psi
\label{cont}
\end{equation}
which describes the motion of a quantum particle of energy $E = k^{2}$
in a random potential $v(x)=k \xi(x)$.
Actually, expression~(\ref{lyap3}) for the inverse localisation length
has long been known to solid state physicists (see, e.g.,
Ref.~\cite{Pas88}) as the high-energy limit of the Lyapunov exponent
for the continuous model~(\ref{cont}).
Thus, the deduction of the inverse localisation length~(\ref{lyap4})
for the Anderson model from expression~(\ref{lyap3}) may be interpreted
as the proof that the continuous model~(\ref{cont}) can be put into
one-to-one correspondence with the discrete lattice~(\ref{andmod})
{\em if, and only if,} the correlation function of the random potential
has the specific form~(\ref{corfun}).
(Obviously, the transposition of results from one model to the other
requires a proper change of the corresponding parameters with
relations like~(\ref{corres}); as a consequence of this swap, the
mathematical correspondence of the two models does not imply an
exact physical equivalence. Models~(\ref{andmod}) and~(\ref{cont}),
for instance, have different unperturbed energy spectra, defined by
the respective dispersion relations $E=2\cos k$ and $E=k^{2}$).

\section{``Mobility edge'' for a stochastic oscillator}
\label{mobsec}

In Ref.~\cite{Izr99} the authors used formula~(\ref{lyap4}) to
investigate the problem of mobility edge for the Anderson
model~(\ref{andmod}). They showed that long-range correlations
in the disorder can generate a continuum of extended electronic
states and they found a way to construct sequences
$\{ \varepsilon_{n} \}$ of site energies giving rise to a Lyapunov
exponent with pre-defined dependence on the energy.
In particular, using this recipe they were able to construct site
potentials that generate a mobility edge even for the 1D
lattice~(\ref{andmod}).

Here, we show how it is possible to solve the analogous problem for the
random oscillator~(\ref{randosc}) taking formula~(\ref{lyap3}) as a
starting point.
More precisely, we will show how to define a continuous noise $\xi(t)$
such that the corresponding Lyapunov exponent $\lambda(\omega)$ has a
pre-definite dependence on the frequency $\omega$. Since the Lyapunov
exponent determines the asymptotic behaviour of the oscillator energy
(we discuss this point more in detail in the next section),
shaping the function $\lambda(\omega)$ through noise control
enables one to determine the energetic behaviour of the oscillator.
In particular, if the noise $\xi(t)$ has the appropriate time
correlations, the corresponding Lyapunov exponent can sharply drop
from positive values to zero when the unperturbed frequency $\omega$
crosses a threshold value. In physical terms that means that
the energetic growth of the oscillator is suppressed when the
frequency reaches a critical value.
The existence of a frequency threshold determining whether the
oscillator is energetically stable or not is the physical counterpart
of a mobility edge, which divides extended states from localised ones
in the Anderson model.
Thus, in spite of the current wisdom that frequency noise produces
energetic instability (see, e.g.,~\cite{Kam92} and references therein),
it turns out that time correlations of the noise may lead to a
suppression of the energy growth. This conclusion follows directly
from the known formula~(\ref{lyap3}), but, to the best of our knowledge,
this implication has not been discussed before in the literature.

To construct a noise $\xi(t)$ that gives rise to a defined Lyapunov
exponent $\lambda(\omega)$, the starting point is the correlation
function $\chi(\tau)$ that can be easily obtained by inverting
formula~(\ref{lyap3})
\begin{displaymath}
\chi(\tau) = \frac{8}{\pi} \int_{-\infty}^{\infty} \lambda(\omega)
e^{2i \omega \tau} \; d \omega .
\end{displaymath}
Once the correlation function $\chi(\tau)$ is known, we can obtain
a stochastic process $\xi(t)$ satisfying the conditions~(\ref{noisprop})
by means of the convolution product
\begin{equation}
\xi(t) = \left( \beta \ast \eta \right) (t) = \int_{-\infty}^{+\infty}
\beta(s) \eta(s+t) \; ds ,
\label{conv}
\end{equation}
where the function $\beta(t)$ is related to the Fourier transform
$\tilde{\chi} (\omega)$ of the noise correlation function through
the formula
\begin{displaymath}
\beta(t) = \int_{-\infty}^{+\infty} \sqrt{\tilde{\chi} (\omega)}
e^{i \omega t} \; \frac{d \omega}{2 \pi}
\end{displaymath}
and $\eta(t)$ is any stochastic process such that
\begin{equation}
\begin{array}{ccc}
\langle \eta(t) \rangle = 0 & \mbox{and} & \langle \eta(t) \eta(t')
\rangle = \delta (t-t') .
\end{array}
\label{eta}
\end{equation}
Formula~(\ref{conv}) defines the family of noises corresponding to a
specific form $\lambda (\omega)$ of the frequency-dependent Lyapunov
exponent and constitutes the solution to the ``inverse problem'' (i.e.,
determination of a noise $\xi(t)$ that generates a pre-defined Lyapunov
exponent).

As an example, we can consider the Lyapunov exponent
\begin{equation}
\lambda (\omega) = \left\{
                   \begin{array}{cc} 1 & \mbox{if} \;\; \;|\omega|<1/2 \\
                                     0 & \mbox{otherwise}
                   \end{array} \right. ,
\label{edge}
\end{equation}
whose frequency dependence implies that the random oscillator undergoes
a sharp transition for  $|\omega|=1/2$, passing from an energetically
stable condition to an unstable one.
Following the described procedure it is easy to see that the Lyapunov
exponent~(\ref{edge}) is generated by a noise of the form
\begin{displaymath}
\xi(t) = \frac{\sqrt{8}}{\pi} \int_{-\infty}^{+\infty} \frac{\sin(s)}{s}
\eta(s+t) \; ds ,
\end{displaymath}
with $\eta(t)$ being any random process with the statistical
properties~(\ref{eta}).

At this point, it is opportune to stress that the mathematical identity
of Eqs.~(\ref{osc}) and~(\ref{cont}) implies that all features of the
random oscillator~(\ref{randosc}) are shared by the solid state
model~(\ref{cont}). Therefore the mathematical results of this Section
not only imply that noise correlations can make the random
oscillator~(\ref{randosc}) stable; they also represent a recipe to
construct a random potential generating a mobility edge for the
model~(\ref{cont}).

\section{Transmission through a disordered barrier}
\label{trasec}

We are now in the position to see how the analogy between the
quantum model~(\ref{andmod}) and the random
oscillator~(\ref{randosc}) can be used not only to compute the
localisation length in the Anderson model but also to deal with
problems both more challenging  and of greater physical interest,
such as the study of the transmission properties of a disordered
barrier. In this Section we show how the random oscillator
formalism allows us to tackle this problem and how it is possible
to obtain expressions for the transmission coefficient as a
function both of the sample length and of the inverse localisation
length~(\ref{lyap3}).

More specifically, let us consider the case of a 1D disordered lattice
of $L$ sites sandwiched between two semi-infinite perfect leads.
Mathematically the problem is defined by the Schr\"{o}dinger
equation~(\ref{andmod}), where the site energies $\varepsilon_{n}$
are now equal to zero for $n < 1$ and $n > L$, while for $1 \leq n \leq
L$ they are assumed to be correlated random variables.
In~\cite{Izr95} it was shown that the transmission coefficient $T_{L}$
through the $L$-sites segment can be expressed in terms of the
classical map~(\ref{map}) as
\begin{equation}
T_{L} = \frac{4}{2 + r_{1}^{2}(L) + r_{2}^{2}(L)} ,
\label{trans1}
\end{equation}
where $r_{1}(L)$ and $r_{2}(L)$ represent the radii at the $L$th
step of the map trajectories starting from the phase-space points
$P_{1} = (x_{0}=1,p_{0}=0)$ and $P_{2} = (x_{0}=0,p_{0}=1)$
respectively.
An analogous formula was given in~\cite{Pas88} for continuous models
like the one defined by Eq.~(\ref{cont}).

Formula~(\ref{trans1}) constitutes the bridge that makes possible
to link the transmission properties of a disordered barrier to the 
time evolution of the energy $r^2$ of the stochastic
oscillator~(\ref{randosc}).
Taking this formula as starting point, one can analytically study the
transport properties of a random barrier in two distinct cases: the
ballistic regime, when the width of the barrier is much less that the
localisation length for the infinite lattice, and the localised regime,
when the vice-versa is true. The two cases are respectively identified
by the conditions $L \ll l_{\infty}$ and $l_{\infty} \ll L$, where we
use the symbol $l_{\infty}$ to denote the inverse of the Lyapunov
exponent~(\ref{lyap4}) and we are assuming that the lattice step is
unitary, so that we can refer to $L$ both as the number of barrier
sites and as the length of the barrier.
We will evaluate the transmission properties first in the ballistic and
then in the localised regime.

Before proceeding to the discussion of the two cases, we observe that our
use of the continuous model~(\ref{randosc}) makes the results of this
Section valid for both continuous models like~(\ref{cont}) and for the
discrete lattice~(\ref{andmod}). The same formulae apply to both cases,
with the localisation length $l_{\infty}$ taking the form~(\ref{lyap3})
or~(\ref{lyap4}) depending on whether the formulae refer to the
continuous or the discrete model.
We also note that the results of this section were obtained long ago
for continuous models (see, e.g.,~\cite{Pas88} and references therein);
what is new here is their application to the discrete case and the
approach used in their derivation, which sets the mathematical results
in a different physical perspective.

\subsection{The ballistic regime}

In the ballistic regime, i.e., when $L \ll l_{\infty}$, one has
$r_{1,2}(L) \simeq 1$ and expression~(\ref{trans1}) can be
written in the form
\begin{equation}
T_{L} = 1 + \frac{2 - r_{1}^{2}(L) - r_{2}^{2}(L)}{4} + \ldots
\label{trans2}
\end{equation}
Another quantity of physical interest is the resistance of the
finite barrier, which is here defined as the inverse of the
transmission coefficient
\begin{equation}
R_{L} = T_{L}^{-1} = \frac{2 + r_{1}^{2}(L) + r_{2}^{2}(L)}{4} .
\label{resis}
\end{equation}

A glance at expressions~(\ref{trans2}) and~(\ref{resis}) reveals that,
in order to obtain the {\em average} value of these physical quantities,
one has to compute the average of the squared radii $r_{1}^{2}(L)$ and
$r_{2}^{2}(L)$ over different disorder realizations.
To achieve this goal, one can rely on the method developed by
Van Kampen to study random oscillators and other stochastic
models~\cite{Kam92, Kam74}.
Van Kampen's approach is based on the construction of a dynamical
equation for the average moments of the position and momentum of the
random oscillator. For the second moments one has
\begin{equation}
\frac{d}{dt} \left( \begin{array}{c} \langle x^{2} \rangle \\
                                     \langle p^{2} \rangle \\
                                     \langle px \rangle
                     \end{array} \right) =
\bf{A} \left( \begin{array}{c} \langle x^{2} \rangle \\
                                     \langle p^{2} \rangle \\
                                     \langle px \rangle
                     \end{array} \right)
\label{momev}
\end{equation}
where the evolution matrix is
\begin{equation}
\bf{A} = \left( \begin{array}{ccc} 0 & 0 & 2 \omega \\
\epsilon_{1} + \epsilon_{2} & - \epsilon_{1} + \epsilon_{2} & -2 \omega \\
- \omega + \epsilon_{3} & \omega & - \epsilon_{1} + \epsilon_{2}
\end{array} \right)
\label{evmat}
\end{equation}
with
\begin{displaymath}
\begin{array}{ccl}
\epsilon_{1} & = & \displaystyle
\int_{0}^{\infty} \chi(\tau) \; d\tau \\
\epsilon_{2} & = & \displaystyle
\int_{0}^{\infty} \chi(\tau) \cos \left( 2 \omega
\tau \right) \; d\tau \\
\epsilon_{3} & = & \displaystyle
\int_{0}^{\infty} \chi(\tau) \sin \left( 2 \omega
\tau \right) \; d\tau .
\end{array}
\end{displaymath}
For the general case of coloured noise, Eq.~(\ref{momev}) is correct up to
order $O(\epsilon) = O(\xi^{2})$; for the special case of white noise,
however, it turns out to be {\em exact}.

One can extract substantial information from Eq.~(\ref{momev}); in
particular, it is possible to obtain the behaviour of the average
squared radii $r_{1}^{2}(t)$ and $r_{2}^{2}(t)$ for $t \rightarrow 0$
\begin{eqnarray*}
\langle r_{1}^{2}(t) \rangle & = & 1 + ( \epsilon_{1} + \epsilon_{2} )
t + o(t^{2}) \\
\langle r_{2}^{2}(t) \rangle & = & 1 + (- \epsilon_{1} + \epsilon_{2} )
t + o(t^{2}) .
\end{eqnarray*}
As a consequence one has
\begin{equation}
\langle \frac{r_{1}^{2}(t) + r_{2}^{2}(t) - 2}{4} \rangle =
\frac{1}{2} \epsilon_{2} t + o(t^{2})
\label{aver}
\end{equation}
Notice that this equations are correct up to order $O(t^{2})$, so that
it is meaningful to retain the distinction between the parameter
$\epsilon_{1}$ and $\epsilon_{2}$.
Using the result~(\ref{aver}) one arrives at the following expressions
for the average transmission coefficient and resistance
\begin{displaymath}
\langle T_{L} \rangle = 1 - 2 \frac{L}{l_{\infty}} + O \left( {\left(
\frac{L}{l_{\infty}} \right)}^{2} \right)
\end{displaymath}
and
\begin{displaymath}
\langle R_{L} \rangle = 1 + 2 \frac{L}{l_{\infty}} + o \left( {\left(
\frac{L}{l_{\infty}} \right)}^{2} \right) .
\end{displaymath}
These formulae show that in the ballistic regime the averages of both
the transmissivity and the resistance are linear functions of the
thickness $L$ of the disordered layer. In addition, the average
resistance coincides with the inverse of the average transmissivity
\begin{displaymath}
\langle T_{L}^{-1} \rangle \simeq {\langle T_{L} \rangle}^{-1} .
\end{displaymath}

\subsection{The localised regime}

In the localised regime the barrier extends over several localisation
lengths: $L \gg l_{\infty}$. In this case, to evaluate the average value
of the transmission coefficient~(\ref{trans1}) it is convenient to
determine the probability distribution for the random variable $r$.
We observe that for $L \gg l_{\infty}$ the radius increases
exponentially and one has
\begin{equation}
r_{1}(L) \simeq r_{2}(L) \simeq r(L),
\label{asympt}
\end{equation}
with probability equal to one, regardless of the initial condition.
As a consequence we can drop the subscripts $1$ and $2$ write the
transmission coefficient in the simplified form
\begin{equation}
\langle T_{L} \rangle \simeq \langle \frac{2}{1 + r^{2}(L)} \rangle .
\label{trans3}
\end{equation}
From the mathematical point of view, the problem of computing the
average~(\ref{trans3}) can be better handled by introducing the
logarithmic variable $z=\ln r$. The dynamics of the random
oscillator~(\ref{randosc}) is then determined by the equations
\begin{equation}
\begin{array}{ccl}
\displaystyle \dot{z} & \displaystyle = &
\displaystyle - \frac{1}{2}  \xi(t) \sin 2 \theta \\
\displaystyle \dot{\theta} & \displaystyle = &
\displaystyle \omega + \xi(t) \sin^{2} \theta .
\end{array}
\label{stochas}
\end{equation}

System~(\ref{stochas}) belongs to the class of stochastic differential
equations of the form
\begin{equation}
\dot{u}_{i} = F^{(0)}_{i}({\bf{u}}) + \alpha F^{(1)}_{i}({\bf{u}},t)
\label{vankam1}
\end{equation}
where $F_{i}^{(0)}({\bf{u}})$ represents a sure function of $\bf{u}$
perturbed by a stochastic function $\alpha F_{i}^{(1)}({\bf{u},t})$
with $\alpha \ll 1$.
Indeed, one can reduce the system~(\ref{stochas}) to the
form~(\ref{vankam1}) by defining the vectors of Eq.~(\ref{vankam1}) as
\begin{displaymath}
\begin{array}{ccc}
{\bf{u}} = \left( \begin{array}{c} z\\
                                  \theta
                  \end{array} \right) , &
{\bf{F}}^{(0)} = \left( \begin{array}{c} 0 \\
                                       \omega
                        \end{array} \right) , &
{\bf{F}}^{(1)} = \left( \begin{array}{c}
                 -\frac{1}{2 \alpha} \xi(t) \sin(2 \theta) \\
                  \frac{1}{\alpha} \xi(t) \sin^{2} \theta
                  \end{array} \right)
\end{array}
\end{displaymath}
with $\alpha = \sqrt{\langle \xi^{2}(t) \rangle}$.
It is known that a stochastic differential equation of the
form~(\ref{vankam1}) can be associated to a partial differential equation
whose solution $P({\bf{u}},t)$ represents the probability distribution
for the random variable $\bf{u}$~\cite{Kam92}.
This partial differential equation can be written as
\begin{equation}
\begin{array}{l}
\displaystyle
\frac{\partial P}{\partial t} = 
- \sum_{i} \frac{\partial}{\partial u_{i}} \left(
F_{i}^{(0)} P({\bf{u}},t) \right) \\
\displaystyle +
\alpha^{2} \sum_{i,j} \frac{\partial}{\partial u_{i}}
\int_{0}^{\infty}
d \tau \, \langle F^{(1)}_{i} ({\bf{u}},t) \frac{d(u^{-\tau})}{d(u)}
\frac{\partial}{\partial u_{j}^{-\tau}} F^{(1)}_{j} ({\bf u}^{-\tau},
t-\tau) \rangle \frac{d(u)}{d(u^{-\tau})} P({\bf{u}},t) \\
+ o \left( \alpha^{2} \right)
\end{array}
\label{vankam2}
\end{equation}
where ${\bf{u}}^{t}$ stands for the flow defined by the deterministic
equation $\dot{{\bf{u}}} = {\bf{F}}^{(0)}({\bf{u}})$,
$d(u^{-\tau})/d(u)$ is the Jacobian of the transformation ${\bf{u}}
\rightarrow {\bf{u}}^{-\tau}$, and the symbol $o \left( \alpha^{2}
\right)$ represents omitted terms of order higher than the second in
the perturbative parameter $\alpha$.
Thus, in the case of weak stochasticity ($\alpha \ll 1$), one can
describe the dynamical behaviour of the system~(\ref{vankam1}) with
an approximate equation of the Fokker-Planck kind.

In the present case, the approximate Fokker-Planck equation~(\ref{vankam2})
associated to the dynamical system~(\ref{stochas}) takes the form
\begin{equation}
\begin{array}{ccl}
\displaystyle
\frac{\partial P}{\partial t} \left( \theta, z, t \right) & = &
\displaystyle
- \omega \frac{\partial P}{\partial \theta} + \frac{1}{4} \sin (2 \theta)
\frac{\partial }{\partial \theta} \left\{ \left[ - \epsilon_{1} +
\epsilon_{2} \cos (2 \theta) + \epsilon_{3} \sin (2 \theta) \right]
\frac{\partial P}{\partial z} \right\}  \\
\displaystyle
& + & \displaystyle
\frac{1}{2} \frac{\partial }{\partial \theta}
\left\{ \sin^{2} (\theta) \left[ \epsilon_{3}
\cos (2 \theta) - \epsilon_{2} \sin (2 \theta) \right]
\frac{\partial P}{\partial z} \right. \\
\displaystyle
& + & \displaystyle
\left. \sin^{2} (\theta)
\frac{\partial }{\partial \theta}
\left[ \left( \epsilon_{1} - \epsilon_{2} \cos (2 \theta) - \epsilon_{3}
\sin (2 \theta) \right) P \right]  \right\} \\
& + & \displaystyle
\frac{1}{4} \sin (2 \theta) \left[ \epsilon_{2} \sin (2 \theta) -
\epsilon_{3} \cos (2 \theta) \right] \frac{\partial^{2} P}{\partial z^{2}} .
\end{array}
\label{fokpl}
\end{equation}
We remark that this equation is correct to the second order in $\xi(t)$
in the general case of coloured noise; in the special case when the noise
$\xi(t)$ is {\em white}, however, it can be shown that
Eq.~(\ref{fokpl}) becomes {\em exact}.


Once we dispose of the Fokker-Planck equation~(\ref{fokpl}) for the
general distribution $P(z,\theta,t)$, we can consider that, in order
to evaluate the average of the transmission coefficient~(\ref{trans1}),
we actually need only the probability distribution for the radial
variable $r$ (or for the equivalent logarithmic variable $z$).
Therefore, we do not have to solve Eq.~(\ref{fokpl}) in all its
generality and we can instead consider the restricted Fokker-Planck
equation 
\begin{eqnarray*}
\frac{\partial}{\partial t} \int_{0}^{2 \pi} P(\theta, z,t) \; d\theta
=  \frac{1}{8} \int_{0}^{2 \pi} \left[ \left( 1 - \cos (4 \theta)
\right) \epsilon_{2} - \sin (4 \theta) \epsilon_{3} \right]
\frac{\partial^{2} P}{\partial z^{2}} \; d\theta \\
+ \frac{1}{4} \int_{0}^{2 \pi} \left[ 2 \epsilon_{1} \cos (2 \theta)
- \epsilon_{2} \left( 1 + \cos (4 \theta) \right) - \epsilon_{3}
\sin (4 \theta) \right] \frac{\partial P}{\partial z} \; d\theta
\end{eqnarray*}
obtained by integrating Eq.~(\ref{fokpl}) over the redundant
angular variable.
To proceed further we assume that, after a short-lived transient, the
probability distribution takes the form
\begin{equation}
P(\theta,z,t) \simeq \frac{1}{2 \pi} P(z,t) .
\label{flat}
\end{equation}
This assumption can be justified on the grounds that, for weak noise,
the dynamics of the angular variable is approximately ruled by the
equation $\dot{\theta} \simeq \omega$. This implies that, after a
sufficiently long time (of the order of some periods $2 \pi/ \omega$),
the angular variable will have swept the whole interval $[0:2 \pi]$ in
a almost uniform way. That makes reasonable to suppose that, for times
$t \gg 2 \pi/ \omega$, the angular distribution is flat (excluding of
course the exceptional case when $\omega \simeq 0$, i.e., when the
energy value lies in a neighbourhood of the band edge).

As a consequence of the hypothesis~(\ref{flat}), one eventually gets
the reduced Fokker-Planck equation for the $z$ variable
\begin{equation}
\frac{\partial P}{\partial z} (z,t) = \lambda \left[ -
\frac{\partial P}{\partial t} (z,t) + \frac{\partial^{2} P}{\partial z^{2}}
(z,t) \right] ,
\label{refokpl}
\end{equation}
where $\lambda$ is the Lyapunov exponent~(\ref{lyap4}).
Eq.~(\ref{refokpl}) has the form of a heat equation with a constant drift;
its solution is therefore
\begin{equation}
P(z,t) = \frac{1}{\sqrt{2 \pi \lambda t}} \exp
-\frac{\left( z - \lambda t \right)^{2}}{2 \lambda t} .
\label{gauss}
\end{equation}
This solution satisfies the initial condition $P(z,t=0) = \delta(z)$,
i.e., we have assumed that at time $t=0$ one has $r=1$, as is the case
for the initial conditions $P_{1}$ and $P_{2}$. The initial condition,
however, is somewhat arbitrary, since the equation~(\ref{refokpl}) is
correct only for times $t \gg 2 \pi/ \omega$.

Knowledge of the distribution~(\ref{gauss}) makes possible to compute
the average transmission coefficient in the localised regime.
Using probability~(\ref{gauss}) we can actually evaluate
expression~(\ref{trans3}) and thus obtain
\begin{equation}
\langle T_{L} \rangle = \int_{-\infty}^{+\infty} \frac{2}{1 + \exp (2z)}
P(z,L) \; dz \simeq \sqrt{\frac{\pi l_{\infty}}{2 L}} \exp \left(
-\frac{L}{2 l_{\infty}} \right) .
\label{trans4}
\end{equation}
As a result, in the limit $L \rightarrow \infty$ one has
\begin{equation}
- \frac{1}{L} \ln \langle T_{L} \rangle = \frac{\lambda}{2}  .
\label{decay}
\end{equation}

Formulae~(\ref{trans4}) and~(\ref{decay}) show that in the localised
regime the transmission coefficient decreases exponentially with the
width of the disordered barrier and they  provide the correct rate of
exponential decay. It must be pointed out, however, that
expression~(\ref{trans4}) fails to reproduce the exact pre-exponential
factor, which actually scales as $(l_{\infty}/L)^{3/2}$ (for
approximation-free results see~\cite{Pas88} and references therein).
This partial shortcoming must be attributed to the two approximations
made in the derivation of formula~(\ref{trans4}), i.e.,
i)~assumption~(\ref{asympt}) that allows the
substitution of the exact expression~(\ref{trans1}) for the
transmission coefficient with the simplified form~(\ref{trans3})
and ii)~hypothesis~(\ref{flat}) about the angular dependency of the
probability distribution $P(\theta,z,t)$.
Both assumptions are admittedly incorrect for very short times,
i.e., for distances $L$ which are small on the length scale defined
by $l_{\infty}$. Thus we are led to the conclusion that in
formula~(\ref{trans4}) the exponential factor is determined by
the long-time behaviour of the random oscillator (which is correctly
described in our approach), while the pre-exponential factor is
strongly influenced by the short-time dynamics of the
oscillator. 
It is interesting to notice that an incorrect  pre-exponential
factor proportional to $(l_{\infty}/L)^{1/2}$ was also obtained
in~\cite{Pas88} studying a continuous solid-state model with a
different approach. In that study, however, the physical meaning of
the adopted simplifying hypotheses was not so transparent as in the
present case, where the analogy between models~(\ref{andmod})
and~(\ref{randosc}) makes possible to gain an intuitive comprehension
of the mathematical approximations.

Beside allowing one to compute the average of the transmission
coefficient, the probability distribution~(\ref{gauss}) makes
possible to determine the average value of other physical
quantities which are relevant for a thorough description of the
transport properties of a disordered barrier. The logarithm of
the transmission coefficient and the resistance~(\ref{resis}) are
standard choices for the complete analysis of the conductance
problem.

The interest for the logarithm of the transmission coefficient
stems from the fact that, unlike the transmission coefficient
itself, the logarithm $\ln T_{L}$ is a {\em self-averaging} variable
and therefore a physically more sound parameter for the definition of
the transport features of the disordered barrier (see,
e.g.,~\cite{Pas88}).
In the present framework, the average of the logarithmic
transmissivity can be computed as follows. First, we observe again
that in the localised regime condition~(\ref{asympt}) is fulfilled
for almost every realization of the disorder so that we can write
\begin{displaymath}
\langle \ln T_{L} \rangle \simeq \langle \ln \frac{2}{1 + r^{2}(L)}
\rangle .
\end{displaymath}
This expression can be put in the equivalent form
\begin{equation}
\langle \ln T_{L} \rangle = - \langle \ln \left( r^{2} \right) \rangle +
\ln(2) - \langle \ln \left( 1 + \frac{1}{r^{2}} \right) \rangle .
\label{lntl1}
\end{equation}
We now observe that for every $x>0$ the logarithm satisfies the
relations $0 < \ln \left( 1 + x \right) < x$; hence the last term
on the r.h.s of the preceding equation must obey
\begin{equation}
0 < \ln \left( 1 + \frac{1}{r^{2}} \right) \rangle < \langle
\frac{1}{r^{2}} \rangle = 1 ,
\label{lntl2}
\end{equation}
where we have made use of distribution~(\ref{gauss}) to evaluate
the average of $1/r^{2}$.
Relations~(\ref{lntl1}) and~(\ref{lntl2}) imply that in the limit
$L \rightarrow \infty$ one has
\begin{displaymath}
- \frac{1}{L} \langle \ln T_{L} \rangle = \frac{2}{L}
\langle \ln r(L) \rangle .
\end{displaymath}
Substituting in the r.h.s. of this equation the average value of
the variable $z=\ln r$ one finally obtains
\begin{displaymath}
- \frac{1}{L} \langle \ln T_{L} \rangle = 2 \lambda
\end{displaymath}
which shows that the average logarithm of the transmission coefficient
decreases linearly with the barrier width in the localised regime.

A third quantity that represents a meaningful statistical
characteristic of the disordered barrier is given by the inverse
of the transmission coefficient, i.e., by the resistance~(\ref{resis}).
As we did in the previous cases, we rely on the condition~(\ref{asympt})
to write the resistance in the form
\begin{equation}
R_{L} \simeq \frac{1}{2} \left( 1 + r^{2}(L) \right) .
\label{resist2}
\end{equation}
Starting from this expression and making use of distribution~(\ref{gauss})
we obtain
\begin{equation}
\langle R_{L} \rangle = \frac{1}{2} \left[ 1 + \exp(\frac{4L}{l_{\infty}})
\right] .
\label{resist3}
\end{equation}

This expression shows that the average value of the resistance increases
exponentially so that the resistance has a multiplicative rather that
additive behaviour as a function of the barrier length. This conclusion
obviously ceases to be valid in the special case in which long-range
correlations of the random potential make the localisation
length $l_{\infty}$ diverge: in this case the disordered
barrier becomes transparent. We underline that, using the recipe
given in Ref.~\cite{Izr99} for the Anderson model -or the prescriptions
of Sec.~\ref{mobsec} for the continuous model~(\ref{cont})- it is possible
to define a random potential such that the corresponding Lyapunov exponent
is zero in certain frequency intervals and positive elsewhere.
As a consequence, the disorder barrier generated by such a potential
will be transparent for electrons with the appropriate energies and
exponentially high otherwise.
This opens the possibility of projecting efficient electronic filters
and agrees with the recent experimental findings discussed
in~\cite{Kuh00}.

As a further consideration, we observe that Eqs.~(\ref{trans4})
and~(\ref{resist3}) show that in the localised regime the inverse of the
average transmission coefficient does {\em not} coincide with the average
of the resistance
\begin{displaymath}
\langle T_{L}^{-1} \rangle \neq {\langle T_{L} \rangle}^{-1}
\end{displaymath}
in contrast to the ballistic regime case.

At this point we wish to remark that the interest of
expression~(\ref{resist3}) goes beyond the definition of the
transport properties of a disordered barrier.
This is so because the resistance $R_{L}$ is strictly
related to the energy $r^{2}$ of the random oscillator~(\ref{randosc}),
as clearly shown by Eq.~(\ref{resist2}).
The exponential increase of the average resistance can therefore
be reinterpretated as energetic instability of the random
oscillator~(\ref{randosc}) on long time scales and 
formula~(\ref{resist3}) can be rewritten in the alternative form
\begin{displaymath}
\langle r^{2} (t) \rangle \propto \exp \left( \gamma_{E} t \right)
\;\;\;\;\; \mbox{for $t \gg 1$} ,
\end{displaymath}
with
\begin{equation}
\gamma_{E} = 4 \lambda
\label{gamm}
\end{equation}
where $\lambda$ is the Lyapunov exponent~(\ref{lyap3}).
This result shows that the energy of the random oscillator grows
exponentially at large times (unless one has $\lambda = 0$) and
that the rate $\gamma_{E}$ of this exponential increase is equal to
{\em four} times the Lyapunov exponent.
We could have computed the energy growth rate also with a different
approach, taking Van Kampen's equation~(\ref{momev}) as a starting point.
In fact, Eq.~(\ref{momev}) determines the time evolution of the
second moments of the position and momentum of the random oscillator;
it is therefore possible to obtain the result~(\ref{gamm}) by
determining the eigenvalue of the evolution matrix~(\ref{evmat}) with
the largest real part.

Incidentally, we observe that some physicists working in the field of
stochastic systems pretend to obtain the rate of orbit divergence
(Lyapunov exponent) by erroneously dividing by a factor {\em two} the
energy growth rate.
The mistake probably stems from (and is equivalent to) the incorrect
assumption that for large times the average of the logarithm of the
energy and the logarithm of the average energy coincide whereas the
real relation is
\begin{displaymath}
\frac{1}{t} \langle \ln r^{2}(t) \rangle = \frac{1}{2t}
\ln \langle r^{2}(t) \rangle
\end{displaymath}
valid in the limit $t \rightarrow \infty$.

As a last remark, we wish to point out another consequence of the
correspondence between the resistance of a disordered barrier and
the energy of the stochastic oscillator~(\ref{randosc}). It is
well known that in the localised regime the resistance $R_{L}$ is
a {\em non-self-averaged} quantity, since the relative fluctuations
of this quantity do not disappear in the macroscopic limit.
Indeed, if we employ the average value~(\ref{resist3}) of the
resistance and use the distribution~(\ref{gauss}) to compute the
average of the square of the resistance~(\ref{resist2}), we obtain
that the root-mean-square deviation of the resistance behaves like
\begin{equation}
\delta R_{L} = \left(
\langle R_{L}^{2} \rangle {\langle R_{L} \rangle}^{-2} - 1
\right)^{1/2} \propto \exp (2L/l_{\infty}) ,
\label{fluct}
\end{equation}
i.e., grows exponentially with the length of the random barrier.
This result is well known to solid-state physicists, but it may
be of some interest to reformulate it in terms of the dynamics of
the stochastic oscillator~(\ref{randosc}). 
Then we can express the meaning of result~(\ref{fluct}) by saying
that the energy of the stochastic oscillator~(\ref{randosc}) is a
quantity whose asymptotic value can fluctuate wildly from one
noise realization to another noise realization. Relative fluctuations
do not die for long times: in this sense it seems that the concept
of non-self-averaging quantity can find an useful application also
in the field of stochastic classical systems.

\section{Conclusions}
\label{concl}

The first part of this paper is devoted to a thorough discussion
of the analogies existing between the Anderson model~(\ref{andmod})
with diagonal correlated disorder and the stochastic
oscillator~(\ref{randosc}) with frequency perturbed by a coloured
noise with correlation function~(\ref{corfun}).
Our analysis shows that the two systems are
equivalent in the sense that there exists a close correspondence
between electronic states on one hand and space trajectories on
the other. Quantitatively, this correspondence manifests itself in
the identity of the inverse localisation length for electronic states
with the exponential rate of divergence of nearby oscillator trajectories.
It is remarkable that this correspondence holds in spite of the fact
that the Anderson model is quantum and discrete (in space) whereas the
random oscillator is classical and continuous (in time).
The analogy between the models~(\ref{andmod}) and~(\ref{randosc})
had already been investigated in~\cite{Tes00} for the basic case of
uncorrelated noise and disorder but the present work extends the
previous conclusions taking into account the effect of {\em
correlations}.

In the second part of the work we discuss some implications of the
parallelism between the models~(\ref{andmod}) and~(\ref{randosc}).
In the first place, we translate the concept of ``mobility edge''
from the field of solid state physics to that of stochastic systems,
showing how time correlations of the frequency noise may produce
energetic stability for the random oscillator~(\ref{randosc}).
We then show how knowledge about the oscillator dynamics on finite
time scales can be used to gain insight about the transport
properties of a finite disordered lattice. This allows us to derive
significant results about electronic transmission in a simple and
physically transparent way. Using the analogy the other way round,
we can also deduce statistical properties of the energy of the
stochastic oscillator from knowledge of the statistical features
of the resistance of a disordered wire.
In passing we also clarify the relation between energetic and orbit
instability for the random oscillator~(\ref{randosc}).

In conclusion, we believe that the bridge built among the fields
of solid-state disordered systems and classical stochastic models
represents a useful way to study the properties of both classes of
systems. The present paper can be considered as an illustration of
how this dual approach works, allowing one to solve old problems by
putting them in a new perspective.

\section{Acknowledgements}

The authors are grateful to N. Makarov and V. Dossetti for fruitful
discussions. L. T. would like to express his appreciation for the support
offered by the Instituto de F\'{\i}sica of Puebla, where much of this
work was done. F. M. I. acknowledges the support by CONACyT (Mexico)
Grant No. 34668-E.

\end{document}